Batch Discovery of New Metal Superhydrides via Chemical Template Theory and Machine Learning

Yuanhui Sun[1,2] and Maosheng Miao[1,*]

[1]*Department of Chemistry and Biochemistry, California State University Northridge, Northridge, CA 91330, USA*

[2]*Suzhou Laboratory, Suzhou 215123, China*

*Correspondence: mmiao@csun.edu

## Abstract

Metal superhydrides, known for their high hydrogen content and polyhedral hydrogen cages, are promising candidates for high-temperature superconductivity. Recent research has emphasized "chemical pre-compression," enabling hydrogen metallization at lower pressures and facilitating the discovery of superconductors with near-room temperature transitions. Despite extensive studies on binary metal superhydrides, there remains a vast, unexplored chemical space, particularly regarding non-integer hydrogen-to-metal ratios. By integrating the "chemical template effect" with machine learning algorithms, we developed a specialized structure discovery workflow that significantly enhances the efficiency of predicting stable superhydrides. Our method led to the identification of 13 new structural prototypes and 31 stable metal superhydrides, representing a 25% increase in discoveries. The 3D hydrogen clathrates in these compounds are strongly correlated with high superconducting transition temperatures, and our approach achieves a remarkable 70% increase. Most of these structures contain over 50 atoms per primitive cell, with the $I4/m$ $M_{10}H_{84}$ prototype having the largest unit cell, containing 94 atoms. Additionally, 19 of the newly identified superhydrides exhibit superconducting transition temperatures ($T_c$) exceeding 100 K, highlighting the potential for higher $T_c$ materials within the 3D hydrogen clathrate structures. The method also shows good potential to search for ternary superhydrides on a large scale.

## Introduction

Metal superhydrides are defined as a class of hydrides with exceptionally high hydrogen content. Many of these compounds feature polyhedral hydrogen (H) cages that encapsulate metal (M) atoms. Recent advances in superhydride research, driven by both experimental and theoretical work, have been inspired by the concept of "chemical pre-compression" [1,2], which enables the metallization of hydrogen at significantly lower pressures than required for elemental hydrogen and has facilitated the discovery of superconductors operating near room temperature [3,4]. The hydrogen lattices within these compounds feature a strong covalent H−H network [5], 3D *s*-orbital conjugation [6], high electron density, and strong electron-phonon coupling, all of which contribute to superconductivity at remarkably high temperatures [7–9]. The first predicted clathrate superhydride, $CaH_6$ [10], was experimentally synthesized with a superconducting critical temperature ($T_c$) of 215 K at 170 GPa [11,12]. Subsequent research efforts to discover superhydrides with higher hydrogen compositions have led to the prediction and/or synthesis of $MH_9$ (M = Y, Ce, Th, Pr, U, Nd, Pa, *etc.*) [13–19], $MH_{10}$ (M = Y, Th, La, Ac, *etc.*) [19,9,20–26], $LaH_{16}$ [27], $MH_{18}$ (M=Y, La, Ac, Ce, Th) [28].

Despite that almost all possible binary metal superhydrides have been studied, and people are forced to search for ternary superhydrides to look for new potential candidate for room temperature superconductors [29–34]; there is still a large chemical space of binary metal superhydrides that has not been systematically explored. Most previous studies have assumed integer $n_H/n_M$ ratios, despite the absence of any chemical rule. Because the interactions between metals and hydrogen lattices are neither covalent nor ionic, their ratios are not restricted by oxidation states. Instead, our recent theoretical findings suggest that the stabilization of metal superhydrides is driven by the "chemical template effect", a mechanism based on electron localization within interstitial regions of the metal lattices that stabilizes the covalent hydrogen lattices [6]. Unlike direct covalent or ionic interactions, this chemical effect is independent of oxidation states and does not impose restrictions on integer $n_H/n_M$ ratios. For instance, the first-known superconducting hydride, $Th_4H_{15}$ (or $ThH_{3.75}$), which exhibits a non-integer ratio of $n$ = 3.75, was discovered as early as 1970 and demonstrated a $T_c$ of 8 K under ambient conditions [35]. More recently, other superhydrides with non-integer $n_H/n_M$ ratios, such as $Eu_8H_{46}$ [36] and $Ba_8H_{46}$ [37], have been successfully predicted and synthesized. However, to date, no systematic exploration of superhydrides with non-integer $n_H/n_M$ ratios has been undertaken.

The primary challenge in incorporating non-integer $n_H/n_M$ ratios into structural research and density functional theory (DFT)-based stability predictions of superhydrides lies in the significant computational costs required. This is primarily due to the larger unit cells involved, along with the dramatically expanded range of possible $n_H/n_M$ ratios when non-integer values are included. Most previous crystal structure prediction (CSP) studies [38–44] optimized the positions of both hydrogen and metal atoms simultaneously, causing the computational complexity to escalate rapidly with increasing structure size. Even for systems with integer $n_H/n_M$ ratios, the number of possible configurations is already substantial. Hydrogen lattices in superhydrides with high $n_H/n_M$ ratios are often depicted as cages, such as the $H_{36}$ cage in $CeH_{18}$ [28] similar to $H_{24}$ cage in $CaH_6$ [10], $H_{29}$ cage in $CeH_9$ [13,14], and $H_{32}$ cage in $LaH_{10}$ [20]. These hydrogen cages consist of weakly interacting $H_n$ clusters and offer far more possible configurations than covalently connected hydrogen lattices.

Our recent investigation into the chemical template interactions between metal and hydrogen lattices introduced an effective approach for discovering new metal superhydrides [6]. Guided by the chemical template theory, we focused on identifying metal lattices with stronger template effects and constructed corresponding superhydrides by introducing a controlled number of hydrogen atoms into the interstitial sites of selected metal lattices. This stepwise approach eliminates many unstable or ineffective configurations caused by weak metal-hydrogen interactions, thereby greatly enhancing the efficiency of structural searches. Building on this understanding, we developed a specialized structure discovery workflow in this study. By establishing a relationship between known metal superhydrides and their thermodynamic stability through machine learning (ML) algorithms, we leveraged the trained ML model to efficiently screen for metal lattices exhibiting strong template effects and employed them in subsequent structural searches. Subsequent high-throughput computational validation identified 13 new structural prototypes and 31 stable metal superhydrides, representing a ~25% increase compared to discoveries over the past decade. Superhydrides containing 3D hydrogen clathrate structures are typically linked to higher superconducting transition temperatures, and this study achieved a ~70% increase in the discovery of such structures. Based on superconducting transition temperature estimates for the newly identified structures, 19 out of the 31 structures exhibit a $T_c$ exceeding 100 K.

## Computational details

**First-principles DFT calculations.** The underlying first-principles DFT calculations were carried out by using the plane-wave pseudopotential method as implemented in Vienna *ab initio* Simulation Package [45,46]. The electron-ion interactions were described by the projector augmented wave pseudopotentials [47,48]. We used the generalized gradient approximation formulated by Perdew, Burke, and Ernzerhof [49] as exchange-correlation functional. A kinetic energy cutoff of 600 eV was adopted for wave-function expansion. The *k*-point meshes with interval smaller than $2\pi \times 0.03$ Å$^{-1}$ for electronic Brillouin zone to ensure that all enthalpy calculations converged within 0.02 eV/atom. The high-throughput first-principles calculations were performed by using the Jilin Artificial-intelligence aided Materials-design Integrated Package (JAMIP), which is an open-source artificial-intelligence-aided data-driven infrastructure designed purposely for computational materials informatics [50]. Detailed information of optimized structures is shown in Table S1, which will not be mentioned in the following discussion.

**ML algorithm.** The prediction of the high-accuracy thermodynamic stability of metal superhydrides was performed using ensembling multiple ML algorithms integrated into the AutoGluon framework, which has demonstrated high reliability in previous works [51–53].

**Structure search.** The structure search starting with metal crystals were carried out by combining a Particle Swarm Optimization (PSO) algorithm [38,39] with first-principles DFT calculations. By evaluating the total energy of these structures, 60% of them with the lowest enthalpies, together with 40% generated new structures, are used to produce the structures of next generation by the structure operators of PSO. All the structure searches stop at 10$^{th}$ generation, where about 400 structures are evaluated in each structure search.

**Phonon spectrum.** Phonon spectra were calculated using a finite-difference supercell approach [54] implemented in the Phonopy code [55]. The lattice constant greater than 10 Å is used as the criterion to set the supercell size, and the default value of 0.01 Å is adopted as the atomic displacement in the supercell.

**Enthalpy of formation.** The enthalpies of formation ($\Delta H_f$) of identified new superhydrides were calculated with respect to a hydride with lower H proportion and $H_2$. We adopted a stable hydride with suitable stoichiometry as the left endpoint of convex hulls to clearly show the $\Delta H_f$ of the new superhydride in each M-H system. Solid $H_2$ in *C2/c* and *Cmca-12* phases [56] are used as the right endpoint in the pressure range of 150–200 GPa and 250–350 GPa, respectively. According to the constructed convex hulls, we can obtain their energy differences relative to the convex hull ($E_{hull}$) at a given pressure. Since the important role of zero-point lattice vibration in determining the stability of superhydrides, we also considered the situation including zero-point energy (ZPE). In addition, the temperature effect (T = 3000 K) is considered here as well using quasiharmonic approximation [57] that introduces volume dependence of phonon frequencies as a part of anharmonic effect as implemented in the Phonopy code [55]. For simplicity, the convex hulls shown in the main text are evaluated without including ZPE at 0 K, while the convex hulls evaluated by including ZPE at 0 K or with including ZPE at 3000 K are shown in supporting information. The detailed $E_{hull}$ values of newly identified superhydrides at different conditions are all listed in Table S2.

## Results and discussion

### Overview of the dedicated structure discovery workflow

To facilitate the exploration of complex compounds with non-integer $n_H/n_M$ ratios, we established a dedicated structure discovery workflow. Instead of simultaneously optimizing the positions of both hydrogen and metal atoms, the workflow first applies machine learning (ML) techniques to identify metal lattices exhibiting significant electron localization at interstitial sites (also called quasi-atoms). Structural searches with varying $n_H/n_M$ ratios are then conducted based on these selected metal lattices to uncover new stable structures. The workflow is schematically illustrated in Figure 1 and comprises six key steps: training set preparation, feature engineering, ML model training, metal lattice candidate preparation, metal lattice screening, and high-throughput computational screening of stable metal superhydrides derived from the selected metal lattices. By employing our structure search approach centered on metal crystals with strong "chemical template" effects, we successfully identified 13 new structural prototypes and 31 new polymorphs of binary metal superhydrides. The stoichiometric ratio *n* in $MH_n$ spans from 5.75 to 13, encompassing both integer and non-integer values. Furthermore, most of these structures contain over 50 atoms in their primitive cells, making the exploration of their structural phase space challenging for conventional methods. In the next, we provide a detailed explanation of the six steps in the workflow.

**Training data preparation.** We first gathered 57 reported metal superhydride prototypes, denoted as $MH_n_Sym_{\mathrm{I}}_P$ in Figure 1a, where $MH_n$ stands for metal superhydrides, $Sym_{\mathrm{I}}$ represents the structural symmetry, and *P* denotes the pressure condition. Given that metals near the *s-d* border tend to form superhydrides with exceptionally high hydrogen contents [6,22], we selected 16 *s-d* border metals for element substitution to expand the structural dataset. Considering that metal superhydrides typically stabilize within the 100–300 GPa pressure range and that structural stability varies significantly under different pressures, we optimized the element-substituted structures at 100, 150, 200, 250, and 300 GPa. This process yielded 4151 optimized and converged structures, which were used to construct the dataset. To evaluate the capability of predicting the thermodynamic stability of metal superhydrides based solely on their metal lattices, we removed the hydrogen atoms and retained the corresponding metal frameworks, denoted as $M[H_n]_0_Sym_{\mathrm{I}}_P$. The thermodynamic stability was assessed by calculating the convex hull energy (see Methods). The dataset was divided into a training set (90%) and a testing set (10%), with both sets showing an overall consistent data distribution, as illustrated in Figure S1.

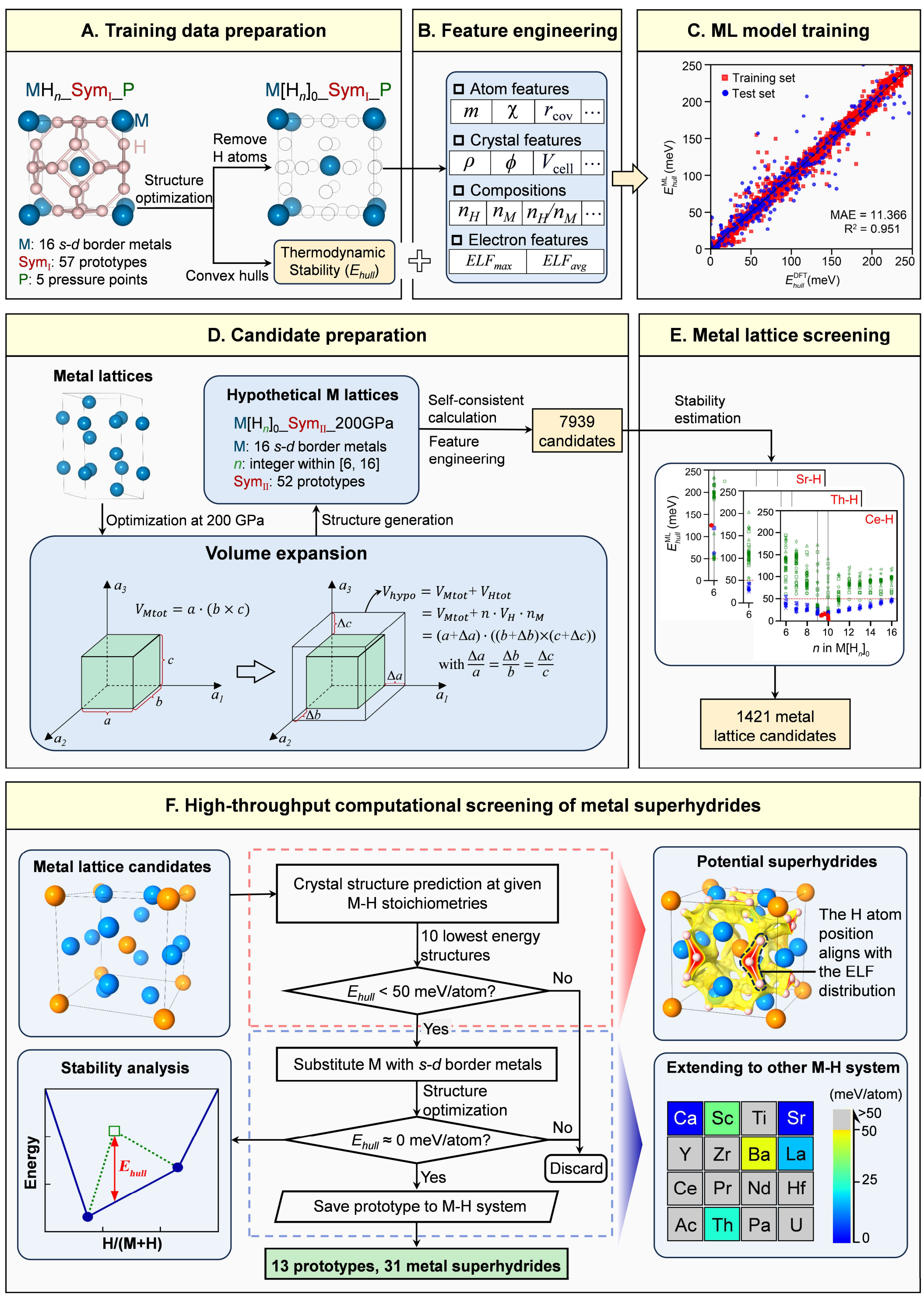

**Figure 1. The designed workflow for discovering metal superhydrides.** (A) Using 57 collected structural prototypes, metal substitution is carried out with *s*-*d* border metals, and structural optimization is performed at various pressure points to obtain the metal lattice and thermodynamic stability ($E_{hull}$) values for the prepared dataset. (B) Atomic features, lattice features, composition, and electronic features of the metal lattices are extracted to create the final feature set for feature engineering, with 46 features used for model training. (C) A robust relationship between the features and thermodynamic stability of superhydrides is established using a stack-ensembling ML strategy based on the AutoGluon framework, achieving an accuracy of MAE = 11.366 meV/atom and $R^2$ = 0.951. (D) The hypothetical configurations of metal lattices after hydrogen atom filling are considered under different $n_H/n_M$ ratios, resulting in 7939 metal lattices for screening. (E) The trained ML model is used to predict the stability of the screened metal lattices. In the case of the Ce-H system, the red scatters represent the newly identified Ce-based superhydrides, along with their corresponding $E_{hull}^{ML}$ values at various H compositions (shown as blue scatters). The black vertical lines indicate the identified stable $CeH_9$ and $CeH_{10}$. (F) Starting from the selected metal lattices, structural searches and high-throughput calculations are employed to identify new stable metal superhydrides.

**Feature engineering.** Constructing the feature set is pivotal for achieving robust predictive performance. In addition to incorporating both elemental and crystal features associated with the metal lattice, we also included electronic characteristics and compositional details of the superhydrides (Figure 1B). The element and crystal features of the metal lattice were directly extracted using the pymatgen [58], Mendeleev [59], and Matminer [60] packages, a common practice in similar studies [61–63]. Given the significant interest of hydrogen composition in the compound, we integrated not only $n_H$ and $n_M$ but also $n_H/n_M$, $n_M/n_H$, $n_H/(n_M + n_H)$, and $n_M/(n_M + n_H)$ into the feature set. Moreover, our previous work revealed a strong correlation between the stability of metal superhydrides and the strength of quasi-atoms within the metal lattices. Accordingly, we introduced two features tied to the lattice's electronic characteristics—namely, the maximum ($ELF_{max}$) and average ($ELF_{avg}$) electron localization function (ELF) values at these interstitials. After the feature engineering step, a total of 46 features were retained to train the model.

**ML model training.** To establish a robust relationship between the features and the thermodynamic stability of superhydrides, we implemented a stack-ensembling ML strategy based on the AutoGluon framework. Compared to a single deep neural network, which is prone to overfitting, this automated stack-ensembling approach optimizes prediction accuracy while ensuring model generalizability, resulting in highly reliable predictions. The model performance for predicting $E_{hull}$ values on the training and test sets is illustrated in Figure 1C. Cross-validation yielded a mean absolute error (MAE) of 11.366 meV/atom, with a coefficient of determination ($R^2$) of 0.951, demonstrating an excellent fit and strong explanatory power regarding the variance of the target variable. According to the feature importance analysis, the top 20 high-weight features are ranked in Figure S2, with detailed descriptions provided in Table S3. The most significant feature is the maximum packing efficiency, which reflects the volume ratio between the metal lattice and its interstitial regions. Since the interstitial volume correlates with the number of hydrogen atoms, this feature can be understood as a combined feature combining properties such as the metal element's radius, valence state, and the number of hydrogen atoms. Additionally, the proposed $ELF_{max}$ and $ELF_{avg}$ parameters ranked 2nd and 7th, respectively, confirming the strong association between the template effect and thermodynamic stability. Two other highly ranked features include the neighbor distance variation and the number of hydrogen atoms in the lattice. The correlation between $E_{hull}$ values and the top nine important features is shown in Figure S3.

**Candidate preparation.** The efficient identification of suitable metal lattices is crucial for advancing the discovery of metal superhydrides. Simple body-centered cubic (BCC), face-centered cubic (FCC), and hexagonal close-packed (HCP) lattices are widely recognized as the typical metal frameworks for common metal superhydrides. However, these high-symmetry structures generally feature only a small number of metal atoms per primitive cell, and few stable superhydrides with non-integer $n_H/n_M$ ratios have been reported (e.g., $Ti_2H_{13}$ [64]). To explore the structural phase space of more complex compounds with non-integer $n_H/n_M$ ratios, we searched for crystal prototypes within the Crystal Lattice Structure sites (http://cst-www.nrl.navy.mil/lattice/) and drew inspiration from variant stacking sequences of close-packed atomic layers [65], ultimately identifying 52 complex metal lattice prototypes. To ensure a comprehensive evaluation, we performed element substitutions on these structural prototypes using 16 *s-d* border metals. Since most known metal superhydrides exhibit stability within the pressure range of 100–300 GPa, all metal lattices were optimized at 200 GPa. Additionally, to better describe the characteristics of metal lattices across various values of *n*, we introduced different amounts of hydrogen into the primitive cells, as shown in Figure 1D, with $n$ ranging from 6 to 16. Notably, the volumetric expansion was limited to cases where the three lattice vectors were scaled proportionally. Naturally, expanding the lattice vectors non-uniformly would generate more structural possibilities, which warrants further exploration. Following self-consistent calculations and feature engineering, we obtained 7939 candidate structures, denoted as $M[H_n]_0_Sym_{\mathrm{II}}_200GPa$.

**Metal lattice screening.** Using the trained ML model, we evaluated the stability of hypothetical superhydrides formed by the 7939 candidate metal lattices. As illustrated in Figure 1E, the green scatters represent the predicted $E_{hull}$ values of various metal lattices at different *n* values within the Ce-H system. The data distribution reveals that most metal lattices exhibit lower $E_{hull}$ values at $n = 9$ and $n = 10$. Notably, the currently reported stable superhydrides in the Ce-H system, $CeH_9$ [13,14]and $CeH_{10}$ [23], align precisely with these regions, as indicated by the black vertical lines. Similar patterns are observed in other M-H systems (Figure S4), providing strong validation for our proposed approach. Additionally, an important trend emerges: metal lattices formed by metals with smaller atomic radii tend to stabilize fewer hydrogen atoms, while those formed by metals with larger atomic radii can accommodate more hydrogen atoms. For example, in the Sc-H and Zr-H systems, stable superhydrides are rarely observed when $n > 10$, whereas in the La-H and Ac-H systems, stable superhydrides are found even at $n = 16$. To identify promising candidates for new stable superhydrides, we established a consensus standard $E_{hull} < 50$ meV/atom [66] as the criterion for structural stability. Consequently, 1421 metal lattices were selected as potential candidates for forming stable superhydrides.

**High-throughput computational screening of metal superhydrides.** Starting with the screened metal lattices, we integrated crystal structure searches with high-throughput computational techniques to accelerate the discovery of new metal superhydrides (Figure 1F). Initially, we conducted a variety of PSO structure searches using the identified metal lattices at the corresponding M-H stoichiometries, as estimated during the volume expansion step, until the $10^{th}$ generation. The 10 lowest-energy structures were then selected, and their $E_{hull}$ values were calculated under the given pressures. Structures with $E_{hull} < 50$ meV/atom were further investigated by substituting the metals with other *s-d* border elements to explore potential stability in different M-H systems. As shown in the upper-right panel of Figure 1F, hydrogen atoms occupy the interstitial sites within the $Pm\overline{3}n$ metal lattice, forming a hydrogen clathrate and leading to a stable metal superhydride. Electronic structure calculations reveal that the positions of the hydrogen atoms correspond closely to the ELF distribution, aligning with the expectations of the template effect. When other *s-d* border metals were substituted, the calculated $E_{hull}$ values (as illustrated in the lower-right panel of Figure 1F) indicate that this structural prototype can be extended to other M-H systems, forming similarly stable configurations.

Finally, we identified 13 new structural prototypes of binary metal superhydrides and 31 polymorphs derived from them. Among these, 9 prototypes (19 polymorphs) exhibit non-integer stoichiometric ratios, while 4 prototypes (12 polymorphs) have integer stoichiometric ratios. Given that ZPE plays a critical role in determining the stability of hydrogen-rich materials, its effect was included in our stability assessments. Moreover, high-temperature conditions, which are typically present during real high-pressure synthesis, can significantly impact the thermal stability of the synthesized compounds. The results, incorporating both ZPE and temperature effects, show that most of the newly discovered structures remain stable or metastable ($E_{hull}$ < 50 meV/atom) (Table S2). The following sections will provide a detailed overview of these structures.

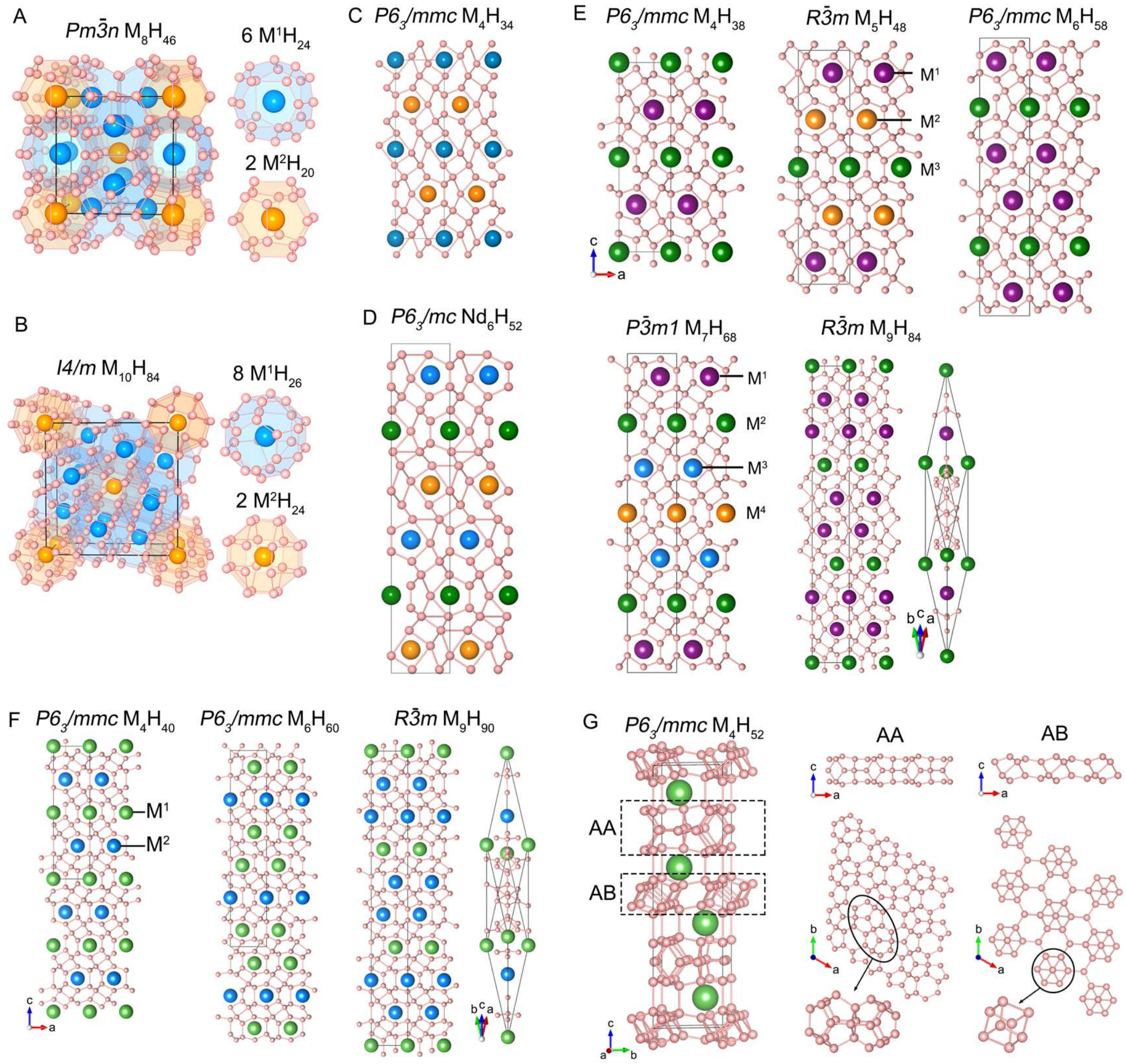

**Figure 2. Crystal structures of newly identified metal superhydrides.** (A) $Pm\bar{3}n$ $M_8H_{46}$ (M= Ca, Sr, Y) contains six $H_{24}$ polyhedral cages with embedded $M^1$ atoms (blue) and two $H_{20}$ polyhedral cages with embedded $M^2$ atoms (orange). (B) $I4/m$ $M_{10}H_{84}$ consists of eight $H_{26}$ polyhedral cages and two $H_{24}$ polyhedral cages. (C) $P6_3/mmc$ $M_4H_{34}$ (M = Pa, Pr). (D) $P6_3/mc$ $M_6H_{52}$ (M = Nd). (E) Five new polymorphs of $MH_{9+\delta}$ ($0 < \delta < 1$, M= Ce, Th), namely *P6₃/mmc* $M_4H_{38}$, $P\bar{3}m1$ $M_5H_{48}$, and *P6₃/mmc* $M_6H_{58}$, $P\bar{3}m1$ $M_7H_{68}$, and $R\bar{3}m$ $M_9H_{84}$. (F) Three new polymorphs of $MH_{10}$ (M = La, Ce, Th), namely *P6₃/mmc* $M_4H_{40}$, *P6₃/mmc* $M_6H_{60}$, and $R\bar{3}m$ $M_9H_{90}$. (G) *P6₃/mmc* $M_4H_{52}$ (M = La, Sr, Ac) consists of an AABB-stacked metal sublattice, where hydrogen atoms between the AA and AB layers form a two-dimensional hydrogen layer.

**Metal superhydrides of $MH_n$ with non-integer stoichiometry *n***

**$Pm\bar{3}n$ $M_8H_{46}$ (M= Ca, Sr, Y).** $Im\bar{3}m$ $CaH_6$, the first clathrate superhydride, was theoretically predicted in 2012 [10] but remained unsynthesized until 2022 [11,12]. In contrast, $P6_3/mmc$ $CeH_9$ [13,14] and $Fm\bar{3}m$ $LaH_{10}$ [20,24] were both theoretically predicted and experimentally synthesized in the same year. Why did it take a decade to synthesize $CaH_6$? Based on our findings, we propose that the newly identified $Pm\bar{3}n$ $Ca_8H_{46}$ (or $CaH_{5.75}$) structure may help elucidate this decay. As shown in Fig. 2A, the $Pm\bar{3}n$ $Ca_8H_{46}$ structure contains six $H_{24}$ polyhedral cages with embedded $Ca^1$ atoms (blue) and two $H_{20}$ polyhedral cages with embedded $Ca^2$ atoms (orange), with 54 atoms per primitive cell. Its $Ca_8$ sublattice is derived from the elemental substitution of $Cr_3Si$ prototype. Notably, the $H_{24}$ and $H_{20}$ cages represent previously unreported polyhedral motifs in metal superhydrides. Each $H_{24}$ cage consists of two hexagons and twelve pentagons, while each $H_{20}$ cage contains twelve pentagons. The relative thermodynamic stabilities of these newly identified metal superhydrides wer assessed using convex hull plots (see methods). A structure lying on the convex hull is thermodynamically stable and cannot decompose into phases with lower formation energy [67]. Without considering $Pm\bar{3}n$ $Ca_8H_{46}$, $Im\bar{3}m$ $CaH_6$ exhibits strong thermodynamic stability at 150 GPa and 200 GPa, as indicated by the convex hull points connected by dashed lines in Fig. 3A. The calculated $\Delta H_f$ is consistent with previous reports [10]. However, when $Ca_8H_{46}$ is included in the analysis, the reconstructed convex hull shows that $CaH_6$ becomes metastable, with $E_{hull}$ values of 25.3 and 21.8 meV/atom at 150 GPa and 200 GPa, respectively. Since hydrogen atoms have high vibrational frequencies, ZPE should be considered to properly evaluate the stability of hydrogen-rich compounds under high pressure. The results indicate that even with ZPE included, $CaH_6$ remains a metastable phase (Figure S5A), with $E_{hull}$ values of 6.0 and 6.5 meV/atom at 150 GPa and 200 GPa, respectively. Moreover, high temperatures typically encountered during high-pressure experiments significantly influence structural stability. At 2000 K, both $Ca_8H_{46}$ and $CaH_6$ lie on the convex hull and are stable at 150 GPa and 200 GPa (Figure S5B). However, $Ca_8H_{46}$ becomes metastable at 3000 K, with $E_{hull}$ values of 8.3 and 3.9 meV/atom at 150 GPa and 200 GPa (Figure S5C), respectively.

Notably, we identified evidence suggesting that $Pm\bar{3}n$ $Ca_8H_{46}$ had already been synthesized, as indicated in previous literature [11]. First, the calculated X-ray diffraction (XRD) pattern of $Pm\bar{3}n$ $Ca_8H_{46}$ closely aligns with the previously unidentified peaks in the experimental synchrotron XRD patterns [11]. Initially, the ambiguous XRD pattern of cell_3 was attributed to the coexistence of $Im\bar{3}m$ $CaH_6$ and $I4/mmm$ $CaH_4$ within the sample, but the *1 and *3 peaks remained unexplained. Our calculations reveal that the XRD pattern of $Pm\bar{3}n$ $Ca_8H_{46}$ corresponds to the three prominent peaks observed in the synchrotron XRD pattern of cell_3 (Figure S6A), providing strong evidence for its presence in the sample. Furthermore, the theoretical equation of state (EOS) of $Pm\bar{3}n$ $Ca_8H_{46}$ shows good agreement with certain experimental EOS measurements (Figure S6B). The calculated EOS of $Im\bar{3}m$ $CaH_6$ is nearly identical to experimental results [11]. Notably, the EOS of $Ca_8H_{46}$ fits well with the experimental data of cell_3 within the pressure range of 120–150 GPa. Certain structures are capable of accommodating hydrogen vacancies and interstitials without significantly compromising their stability. During our search for Ca superhydrides near the $CaH_6$ composition, we identified an $Ia\bar{3}d$ $Ca_8H_{43}$ structure (Table S1) lying near the convex hull and sharing the same Ca sublattice as $Pm\bar{3}n$ $Ca_8H_{46}$. The XRD pattern of $Ia\bar{3}d$ $Ca_8H_{43}$ also shows good agreement with the three prominent peaks in the synchrotron XRD pattern of cell_3 (Figure S6A). The delayed experimental synthesis of $Im\bar{3}m$ $CaH_6$ may be attributed to the presence of highly stable low-energy superhydrides ($Pm\bar{3}n$ $Ca_8H_{46}$ or $Ia\bar{3}d$ $Ca_8H_{43}$) near the $CaH_6$ composition, which remain stable at low temperatures (< 2000 K), posing significant challenges for its synthesis under high-pressure conditions.

By substituting Ca atoms with other *s-d* border metals, we demonstrate that $Pm\bar{3}n$ $Sr_8H_{46}$ is thermodynamically stable at 300 GPa (Figure 3B), and $Pm\bar{3}n$ $Y_8H_{46}$ exhibits metastability at 200 GPa ($E_{hull}$ = 42.9 meV/atom, Figure S5F). $Pm\bar{3}n$ $Sr_8H_{46}$ remains on the convex hull, whether only the ZPE is considered or both ZPE and temperature effects are included (Figures S5D-E). In contrast, the $E_{hull}$ of $Pm\bar{3}n$ $Y_8H_{46}$ remains constant under ZPE-only conditions but drops to 20 meV/atom when both ZPE and a temperature of 3000 K are included (Figures S5G-H). The calculated evolution of $\Delta H_f$ suggests that elevated temperatures further promote their stability. Additionally, the dynamic stabilities of $Pm\bar{3}n$ $M_8H_{46}$ (M = Ca, Sr, Y) are confirmed by the absence of imaginary phonon modes (Figure S7). The $Pm\bar{3}n$ $M_8H_{46}$ structure has been known in a family of clathrate compounds, including $Ba_8Ge_{46}$, $Sr_8Si_{46}$, and several others. More recently, various metal superhydrides, $M_8H_{46}$ (M = Lu, Ba, La, or Eu), with the same clathrate structure have been theoretically predicted and/or experimentally synthesized [36,37,68,69].

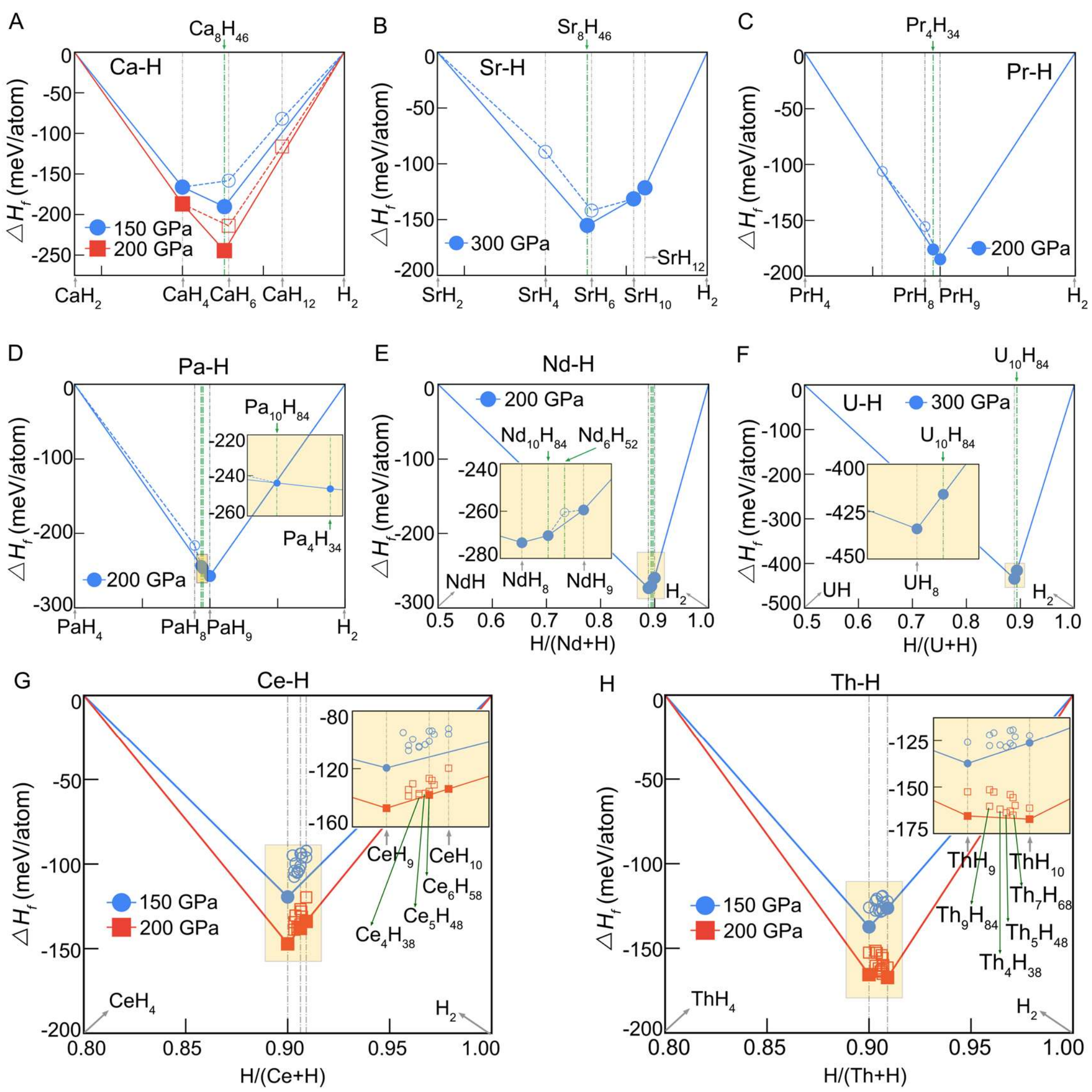

**Figure 3. Estimation of thermodynamic stabilities of newly identified metal superhydrides with non-integer $n_H/n_M$ ratios by employing the convex hull plots at given pressures.** The formation enthalpies of (A) $Pm\bar{3}n$ $Ca_8H_{46}$ in Ca-H system at 150 and 200 GPa, (B) $Pm\bar{3}n$ $Sr_8H_{46}$ in Sr-H system at 300 GPa, (C) $P6_3/mmc$ $Pr_4H_{34}$ in Pr-H system at 200 GPa, (D) $P6_3/mmc$ $Pa_4H_{34}$ and $I4/m$ $Pa_{10}H_{84}$ in Pa-H system at 200 GPa, (E) $I4/m$ $Nd_{10}H_{84}$ and $P6_3/mc$ $Nd_6H_{52}$ in Nd-H system at 200 GPa, (F) $I4/m$ $U_{10}H_{84}$ in U-H system at 300 GPa, (G) several $CeH_{9+\delta}$ $(0 < \delta < 1)$ in Ce-H system at 150 and 200 GPa, and (H) several $ThH_{9+\delta}$ $(0 < \delta < 1)$ in Th-H system at 150 and 200 GPa, respectively. The hollow symbols connected by dashed lines represent the formation enthalpies of previously discovered structures before the inclusion of newly identified superhydrides. The solid points indicate the updated convex hull plots after adding the newly identified

superhydrides. The green vertical dashed lines in these figures show the positions of the newly discovered superhydrides. To clearly highlight the formation energies of these metal superhydrides, the regions marked yellow in Figures 3D-H are zoomed in for better visibility.

**$I4/m$ $M_{10}H_{84}$ (M = Pa, Nd, U).** The primitive cell of $I4/m$ $M_{10}H_{84}$ contains 94 atoms, making it the largest structural prototype identified in our study. This structure consists of eight $H_{26}$ polyhedral cages and two $H_{24}$ polyhedral cages (Figure 2B). The structural parameters of $I4/m$ $M_{10}H_{84}$ (M = Pa, Nd, U) at 200 GPa are listed in Table S1. The $M_8^1M_2^2$ lattice shares a similar metallic framework with the recently reported $I4/m$ $La_4NdH_{50}$ structure [34]. This similarity indicates that such structural prototypes could play a crucial role in the discovery of ternary or multicomponent superhydrides, particularly when the metal sublattice involves multiple inequivalent metal elements. The newly identified $H_{26}$ cage contains squares, pentagons, and hexagons, whereas the $H_{24}$ cage has been reported in $I4/mmm$ $AcH_{12}$ [22]. As shown in Figures 3D-F, the calculated $\Delta H_f$ show that $I4/m$ $M_{10}H_{84}$ (M = Pa, Nd, U) is stable in each M-H system at given pressures. $I4/m$ $U_{10}H_{84}$ remains thermodynamically stable on the convex hull at 300 GPa, whether only the ZPE is considered or both the ZPE and temperature effects are included (Figures S8C, S8F). $I4/m$ $Pa_{10}H_{84}$ becomes metastable with ZPE included ($E_{hull}$ = 22.1 meV/atom, Figure S8A) and exhibits further instability when temperature effects are considered ($E_{hull}$ = 97.1 meV/atom, Figure S8D). However, $I4/m$ $Nd_{10}H_{84}$ becomes metastable when ZPE is considered ($E_{hull}$ = 8.5 meV/atom) but stabilizes on the convex hull when ZPE and temperature effects are both included (Figures S8B, S8E). Additionally, their dynamic stability is further evidenced by the absence of imaginary phonon mode at given pressures (Figure S9).

**$P6_3/mmc$ $M_4H_{34}$ (M = Pa, Pr).** In the Pa-H system, we identified another $P6_3/mmc$ $Pa_4H_{34}$ structure with an $n_H/n_M$ ratio of 8.5, closely approximating the 8.4 found in $I4/m$ $Pa_{10}H_{84}$. The primitive cell of $P6_3/mmc$ $Pa_4H_{34}$ contains 38 atoms, featuring a $Pa_4$ sublattice with double hcp (dhcp) symmetry and a clathrate arrangement of hydrogen atoms (Figure 2C). By substituting Pa atoms with other *s-d* border metals, $P6_3/mmc$ $Pr_4H_{34}$ also remain stable at 200 GPa (Figure 3C). When both ZPE and temperature effects are considered, they retain thermodynamic stability (Figure S10), whereas only $I4/m$ $Pa_{10}H_{84}$ exhibits an $E_{hull}$ of 1.8 meV/atom under combined ZPE and 3000 K conditions. In addition, their dynamic stability is further evidenced by the absence of imaginary phonon mode at given pressures (Figure S11).

**$P6_3/mc$ $M_6H_{52}$ (M = Nd).** $P6_3/mc$ $Nd_6H_{52}$ features a 58 atoms primitive cell with an $n_H/n_M$ ratio of 8.67. The $Nd_6$ sublattice adopts a 6H polytype analogous to that found in SiC [70], specifically following an ABCACB stacking sequence [65], while the hydrogen atoms retain a clathrate arrangement (Figure 2D). As illustrated in Figure 3E, $P6_3/mc$ $Nd_6H_{52}$ exhibits an $E_{hull}$ of 4.8 meV/atom at 200 GPa. Given the magnetic properties of Nd-H systems under high pressure [16], we also evaluated the thermodynamic stability of $P6_3/mc$ $Nd_6H_{52}$ in the ferromagnetic state. Our calculations indicate that this structure becomes stable at 200 GPa (Figure S12). Unfortunately, phonon calculations reveal numerous imaginary frequencies in both the nonmagnetic and ferromagnetic phonon spectra of this structure. Owing to the complexity of magnetic systems, we have not pursued a further investigation, yet we include this structure in our discussion to inform future studies.

**New polymorphs of $\mathrm{MH}_{9+\delta}$ ($0 < \delta < 1$, M= Ce, Th).** In addition to the aforementioned non-integer-ratio metal superhydrides, we identified a series of new polymorphs of $\mathrm{MH}_{9+\delta}$ ($0 < \delta < 1$, M= Ce, Th). As illustrated in Figures 3G and 3H, these structures exhibit extremely low $E_{hull}$ at 150 GPa and 200 GPa. For the Ce-H system, three of the eleven metastable clathrate superhydrides with $E_{hull}$ < 5 meV/atom—*P6₃/mmc* $Ce_4H_{38}$, $P\bar{3}m1$ $Ce_5H_{48}$, and $P6_3/mmc$ $Ce_6H_{58}$, containing 42, 53, and 64 atoms per primitive cell, respectively—were selected for further investigation (Figure 3G). For the Th-H system, three of the eleven metastable clathrate superhydrides with $E_{hull}$ < 5 meV/atom—$P6_3/mmc$ $Th_4H_{38}$, $P\bar{3}m1$ $Th_5H_{48}$, $P\bar{3}m1$ $Th_7H_{68}$, and $R\bar{3}m$ $Th_9H_{84}$, whose primitive cells consist of 42, 53, 75, and 31 atoms, respectively—were similarly investigated (Figure 3H). Their atomic structures are shown in Figure 2E. Their metal sublattice design draws inspiration from the various close-packed sequences of Au [65], with $P\bar{3}m1$ $Th_7H_{68}$ specifically capable of incorporating four inequivalent metal sites. Moreover, their hydrogen lattices incorporate structural characteristics reminiscent of those arising from various combinations of $Fm\bar{3}m$ $LaH_{10}$ and $P6_3/mmc$ $CeH_9$. Upon considering ZPE and temperature effects, the estimated $E_{hull}$ values exhibit minimal changes, with several of these structures stabilizing on the convex hulls (Figure S13). In addition, their dynamic stability is corroborated by the absence of imaginary phonon modes, as shown in Figure S14.

## Metal superhydrides of $MH_n$ with integer stoichiometry *n*

**Three new polymorphs of $MH_{10}$ (M = La, Ce, Th).** One of the most representative binary metal superhydrides to date is $Fm\overline{3}m$ $LaH_{10}$, originally predicted theoretically [20] and subsequently synthesized experimentally [24,71]. Its primitive cell features a La sublattice with FCC symmetry, whereas the primitive cell contains only 11 atoms. To evaluate the workflow proposed in this work, we specifically focused on the $n_H/n_{La} = 10$ composition. As a result, we identified three new polymorphs of $LaH_{10}$ (Figures 2F)—*P6₃/mmc* $La_4H_{40}$, *P6₃/mmc* $La_6H_{60}$, and $R\overline{3}m$ $La_9H_{90}$—whose primitive cells comprise 44, 66, and 33 atoms, respectively. Notably, the $La_4$ sublattice in *P6₃/mmc* $La_4H_{40}$ is in dhcp symmetry, while the $La_9$ sublattice in $R\overline{3}m$ $La_9H_{90}$ corresponds to a Sm-type crystal frequently observed among Lanthanides in the Materials Project [72]. Meanwhile, their hydrogen sublattices consist of $H_{32}$ cages exhibiting various distortions compared to the highly symmetric arrangement found in $Fm\overline{3}m$ $LaH_{10}$. DFT calculations indicate that three newly identified polymorphs are energies comparable to those of $Fm\overline{3}m$ $LaH_{10}$ over the pressure range of 150–300 GPa (Figure 4A). Furthermore, upon substituting La atoms in these polymorphs with other *s-d* border metals, the resulting structures remain energetically favorable in both the Ce-H and Th-H systems. Their computed energies are similar to those of the previously identified $Fm\overline{3}m$ $CeH_{10}$ and $Fm\overline{3}m$ $ThH_{10}$ within 100–200 GPa (Figures 4B-C). When ZPE and temperature effects are considered, the energy differences between the new structures and their corresponding $Fm\overline{3}m$ phases remain within -5 to 20 meV/atom (Figures S15), indicating that they continue to be competitive phases. In addition, their dynamic stability is further affirmed by the absence of imaginary phonon modes, as illustrated in Figure S16.

***P6₃/mmc*** **$M_4H_{52}$ (M = La, Sr, Ac).** To the best of our knowledge, $R\overline{3}m$ $YH_{13}$ is the only reported metastable superhydride with an $n_H/n_M$ ratio of 13, whose primitive cell comprises a simple hexagonal Y lattice surrounded by 13 H atoms in a clathrate arrangement [19]. In this work, we report a new polymorph of $M_4H_{52}$ (M = La, Sr, Ac) with *P6₃/mmc* symmetry, containing 56 atoms per primitive cell (left panel of Figure 2G). The centrosymmetric $M_4$ sublattice in *P6₃/mmc* $M_4H_{52}$ is stacked in AABB sequence. Consequently, the H distribution within the lattice can be categorized into two distinct regions: the AA and AB regions. The H distribution within the AA regions consists of repeating $H_{24}$ clusters (middle panel of Figure 2G), resembling the building unit of the hydrogen lattice in *P6/mmm* $LaH_{16}$ [27]. Notably, this building unit was not addressed by the authors and was first introduced in our recent work [6]. In contrast, the AB region forms a two-dimensional hydrogen layer exhibiting threefold rotational symmetry, constructed from interconnected $H_8$ cubes (right panel of Figure 2G). The distorted $H_8$ cubes are frequently observed in *P6₃/mmc* $MH_9$ (M=Ce, Th, etc.) [13,19,26] and $Fm\bar{4}3m$ $MH_9$ (M = Pr and Pa) [15,17,19]. As illustrated in Figures 4D-F, the calculated $\Delta H_f$ shows that *P6₃/mmc* $M_4H_{52}$ (M = La, Sr, Ac) is stable within their respective M-H systems under given pressures. Under the effect of ZPE, these structures remain stable on the convex hull. However, when both ZPE and temperature effects are considered (Figure S17), *P6₃/mmc* $Ac_4H_{52}$ continues to remain stable on the convex hull, whereas *P6₃/mmc* $La_4H_{52}$ and *P6₃/mmc* $Sr_4H_{52}$ become metastable, with convex hull energies of 30.8 and 7.1 meV/atom at 300 GPa, respectively. The analysis suggests that *P6₃/mmc* $M_4H_{52}$ tends to stabilize at low temperatures. Furthermore, their dynamical stabilities are confirmed by the absence of imaginary phonon modes, as shown in Figure S18.

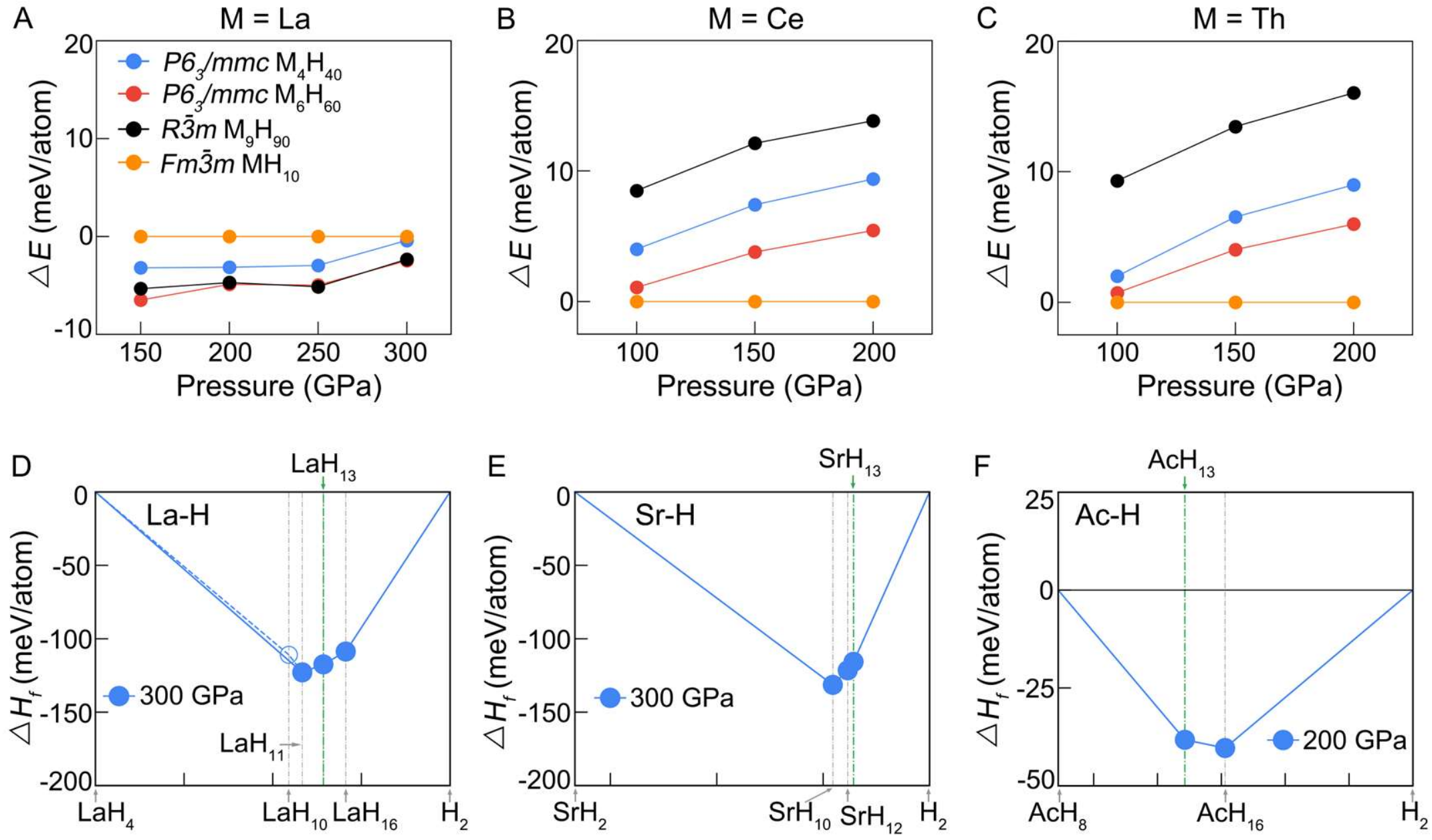

**Figure 4. Estimation of thermodynamic stabilities of newly identified metal superhydrides with integer $n_H/n_M$ ratios at given pressures.** (A-C) Energy differences between new polymorphs of $MH_{10}$ and known $Fm\bar{3}m$ $MH_{10}$ in M-H system (M = La, Ce, Th) at given pressures. (D-F) The formation enthalpies of *P6₃/mmc* $M_4H_{52}$ (M = La, Sr, Ac) in corresponding M-H system at given pressures.

## Superconductivity of newly identified metal superhydrides

A key feature of metal superhydrides is their remarkable superconducting $T_c$. To facilitate rapid estimation of $T_c$ for newly identified structures, we employed the method proposed by Belli et al. [5], which evaluates key hydride-specific characteristics, including 3D connectivity of electron distribution, the hydrogen fraction of the compound, and the hydrogen fraction of the total density of states at the Fermi energy. The reliability of this method has been confirmed in our previous work [6]. The distribution of $T_c$ values for the 31 newly discovered structures as a function of the hydrogen fraction is presented in Figure 5, with detailed values listed in Table S4. Notably, 19 of these structures (~61%) exhibit $T_c$ values exceeding 100 K. These high-$T_c$ structures consist of metal superhydrides formed by La, Ce, and Th, closely aligning with the distribution observed in known superconducting metal superhydrides. However, the accuracy of this estimation method is limited to approximately 60 K [5], making it suitable only for preliminary assessments. More precise evaluations require advanced approaches, such as solving the McMillan-Allen-Dynes formula [73–75]. Given the strong evidence supporting the synthesis of $Pm\bar{3}n$ $Ca_8H_{46}$, we evaluated its superconducting properties at 200 GPa. As shown in Figure 5B, the low-frequency phonons (<20 THz) primarily originate from Ca atoms, while the high-frequency phonons (>20 THz) are predominantly attributed to H atoms. The calculated $T_c$ is 84 K, with a logarithmic average frequency of $\omega_{log}$ = 1058 K and an electron-phonon coupling (EPC) constant of λ = 1.09. The calculated Eliashberg $\alpha^2F(\omega)$ function and the EPC parameter $\lambda(\omega)$ reveal that 24% of λ is contributed by low-frequency phonons, while 76% is derived from high-frequency phonons. The superconducting properties of other newly discovered structures require further in-depth investigation.

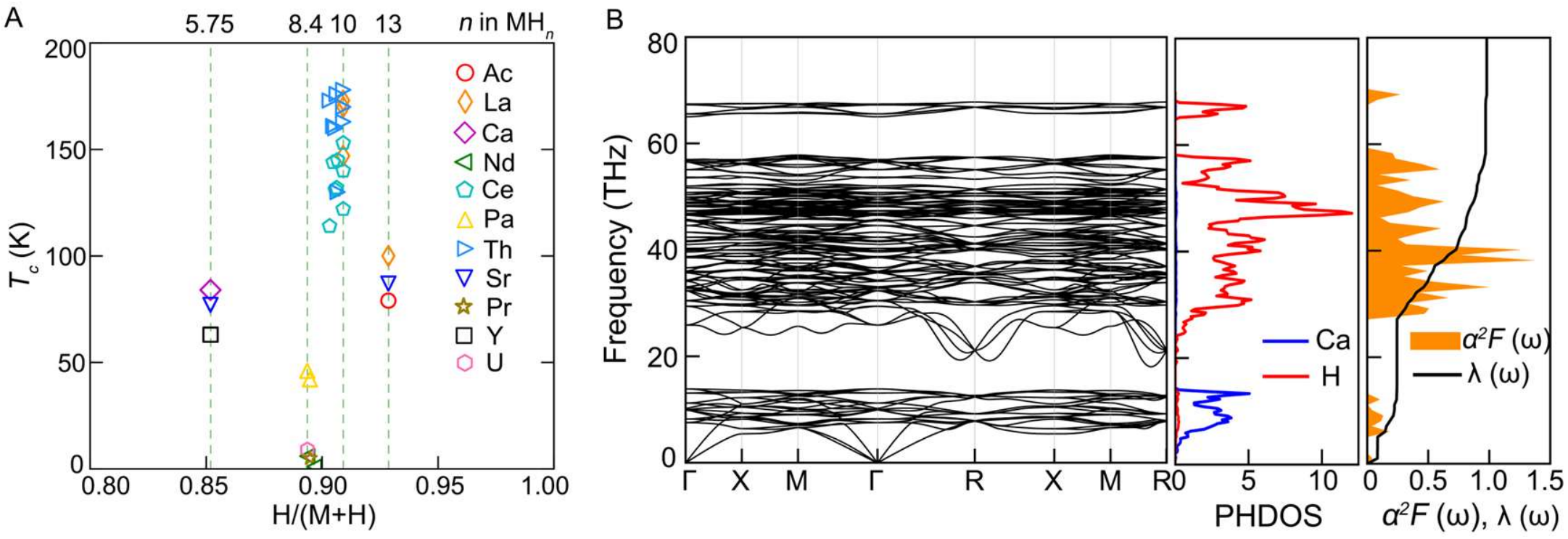

**Figure 5. Superconductivity of newly identified metal superhydrides.** (A) The estimated $T_c$ values of the newly identified metal superhydrides as a function of hydrogen

fraction. Metal superhydrides with the same metal elements are represented by the same symbols. (B) The phonon dispersion curves, PHDOS, Eliashberg $\alpha^2F(\omega)$ function and the EPC parameter $\lambda(\omega)$ of $Pm\overline{3}n$ $Ca_8H_{46}$ at 200 GPa.

## Discussion

In summary, this study establishes a dedicated workflow for discovering metal superhydrides by expanding a dataset derived from 57 structural prototypes identified over the past decade. Through this workflow, we report the discovery of 31 new metal superhydrides and 13 new structural prototypes. Almost all the newly identified structures maintain stability or metastability under the consideration of ZPE and temperature effects. Moreover, phonon calculations confirm their lattice dynamical stability, revealing that these structures exhibit no imaginary frequencies. In terms of the discovery of new structural prototypes, this work, based on 57 previously identified prototypes, has discovered 13 new structural prototypes, achieving an approximately 25% increase in the total number of prototypes in a single effort (Figure 6). Moreover, the 3D hydrogen clathrates in metal superhydrides are often correlated with elevated superconducting transition temperatures, the discovery of clathrate metal superhydrides is therefore critical for the identification of high-temperature superconductors. By building on 17 previously discovered clathrate prototypes, this work identifies 12 additional clathrate prototypes, representing a remarkable ~70% increase. Most of these structures contain over 50 atoms per primitive cell, with the $I4/m$ $M_{10}H_{84}$ prototype having the largest unit cell, containing 94 atoms. Exploring such a vast and complex structural phase space would be nearly impossible without the essential guidance of “chemical templates”. Finally, we provide a rough estimation of the superconducting transition temperatures for the newly identified metal superhydrides, showing that 19 (~61%) of the structures exhibit $T_c$ values exceeding 100 K. These discoveries significantly expand the pool of potential candidates for room-temperature superconducting superhydrides. We encourage researchers interested in synthesis to devote efforts toward experimental validation and exploration in this area.

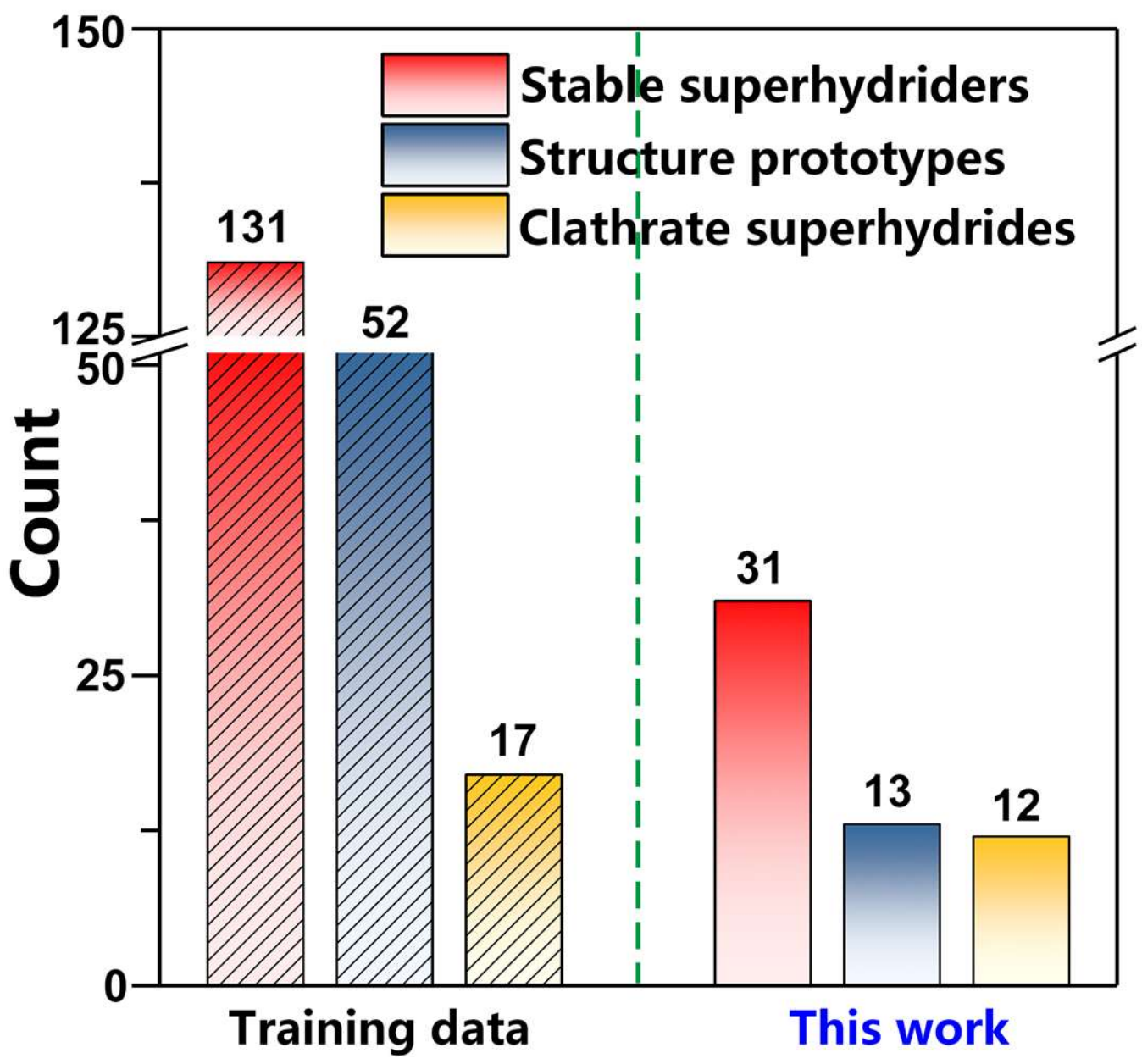

**Figure 6. A comparison of newly discovered and known metal superhydrides.** The left and right panels show the statistics of the training set and the newly discovered superhydrides, respectively. The red, blue, and yellow bars represent the total number of stable superhydrides, the number of structure prototypes, and the number of superhydrides with hydrogen clathrates, respectively.

After exploring many metal superhydrides, we found two general rules of metal lattices that govern the formation of superhydrides. In the first rule, as has been noticed in previous studies [6,22], the metals around the *s*-*d* border form strong templates, therefore, should be focused on the identification of superhydrides. The second rule is also a natural result of the template effect, *i.e.* the metal lattices with a strong tendency to form superhydrides usually have high symmetry. Although low symmetry can enhance electron localizations in some areas of the metal lattice, it will also reduce them in other areas, weakening the template effect overall. As a matter of fact, most superhydrides consist of metal lattices in the most common and simple structures such as BCC, FCC, HCP, simple hexagonal (SH) or structures slightly deformed from them. Although we do not strictly limit our study to high-symmetry lattices, most of the lattices selected based on their strength of template effect are highly symmetric.

As illustrated in Figure 2, these structures contain multiple inequivalent metal atoms, which are intentionally represented in different colors. Unlike the typical BCC, FCC, and HCP metal lattices in known metal superhydrides—each generally containing a single inequivalent metal site—the 31 newly discovered structures usually feature two inequivalent metal sites, with $P\overline{3}m1$ $Th_7H_{68}$ accommodating as many as four. This finding demonstrates that even more complex or lower-symmetrical metal lattices can stabilize hydrogen clathrates. The presence of inequivalent metal positions implies that the surrounding hydrogen environments are different, indicating that these sites may be more favorably occupied by distinct metals. Consequently, substituting elements within these newly discovered structural prototypes could yield ternary bimetallic superhydrides and further multicomponent metal superhydrides, which is an emerging focus in high-temperature superconductivity research. Recent work supports this perspective: the $I4/m$ $La_4NdH_{50}$ structure reported by Semenok et al. [34] shares the same metal lattice prototype as our $I4/m$ $Pa_{10}H_{84}$, and the $La_4Nd$ lattice can be directly obtained from the $Pa_8^1Pa_2^2$ lattice via elemental substitutions. Although their $n_H/n_M$ ratios differ, we believe that the exploration of ternary superhydride systems starting from bimetallic lattices will uncover new stable structures. We anticipate that by screening numerous candidate structures to identify bimetallic lattices with strong template effects and utilizing them for structural searches, followed by high-throughput computational stability verification, many stable ternary bimetallic superhydrides can be discovered.